\documentclass[prl,showpacs]{revtex4}

\usepackage{amsmath,amsfonts}
\usepackage[dvips]{graphicx}

\begin{document}

\title{Cosmological model with energy transfer} 
\author{Marek Szyd{\l}owski}
\email{uoszydlo@cyf-kr.edu.pl}
\affiliation{Complex Systems Research Centre, Jagiellonian University, 
Reymonta 4, 30-059 Krak{\'o}w, Poland}

\begin{abstract}
The observations of SNIa suggest that we live in the acceleration epoch when 
the densities of the cosmological constant term and matter are almost equal. 
This leads to the cosmic coincidence conundrum. As the explanation for this 
problem we propose the FRW model with dark matter and dark energy which 
interact each other exchanging energy. We show that the cubic correction 
to the Hubble law, measured by distant supernovae type Ia, probes this 
interaction. We demonstrate that influences between nonrelativistic matter 
and vacuum sectors are controlled by third and higher derivatives of the scale 
factor. As an example we consider flat decaying $\Lambda(t)$ FRW cosmologies. 
We point out the possibility of measure of the energy transfer by the cubic 
and higher corrections to Hubble's law. The statistical analysis of SNIa data 
is used as an evidence of energy transfer. We find that there were the transfer from the dark energy sector to 
the dark matter one without any assumption about physics governing this 
process. We confront this hypothesis about the transfer with SNIa observations 
and find that the transfer the phantom and matter sector is admissible for 
$\Omega_{\text{m},0}=0.27$. We also demonstrate that it is possible to 
differentiate between the energy transfer model and the variable coefficient 
equation of state model.
\end{abstract}


\maketitle

\section{Introduction}
The modern cosmology, especially observational cosmology, reminds empirical 
science in a crisis phase \cite{Kuhn:1962}. The recent observations of 
supernovae type Ia (SNIa) indicates that the Universe is accelerating at the 
present epoch \cite{Perlmutter:1998np,Riess:1998cb}. We accept that the present 
evolution is well described by the general relativity theory with the 
Robertson-Walker type of space symmetry and the source of gravity is perfect 
fluid then the acceleration of Universe expansion can be explained only in the 
following way. The Universe is filled additionally to nonrelativistic matter 
with dark energy of unknown origin, which violates the strong energy condition. 
When the perfect fluid with the energy $\rho$ and pressure $p$---the source of 
gravity---satisfies the strong energy condition then the explanation of SNIa 
observations requires the modification of the Einstein equations. If we 
postulate the Robertson-Walker symmetry then there are some propositions of 
modification of the Friedmann first integral. Freese and Lewis considered flat 
cosmological models with the additional term $B\rho^n$, where $B$ and $n$ are 
constants \cite{Freese:2002sq}. The parameters of this model were confronted 
with the observation of distant type Ia supernova 
\cite{Zhu:2003sq,Godlowski:2003pd}.

Both approaches can be tested statistically by searching model parameters 
which best fits to the SNIa data. But to distinguish among the models it is 
necessary to take additional observational constraints. When we assume the 
matter density value which is indicated by cosmic microwave background (CMB) 
and galactic counting observations then the most promising model is the dark 
energy model with the cosmological constant. The cosmological constant has 
very long history and is still the source of various problems and troubles 
\cite{Padmanabhan:2002ji,Padmanabhan:2002vv}. The main problem with today face 
of the cosmological constant is that its value is negligible in comparison to 
the Planck mass. In other words the $\Lambda$ cold dark matter ($\Lambda$CDM) 
model looks like the effective theory which gives us the description of the 
phenomenon of acceleration without giving any understanding.

Another problem is why the energy densities of dark energy and of dust-like 
matter are of the same order of magnitude at the present epoch 
\cite{Sahni:2002kh,Gorini:2004by}. This problem is known as a ``cosmic 
coincidence conundrum''. The beginning of the vacuum energy and the 
dust-like matter is related to different epochs separated by very long 
interval of time. One of the possibility of the explanation of this coincidence 
is the intrinsic feedback between the energy density of dark matter and dark 
energy modelled by quintessence scalar fields \cite{Ziaeepour:2003qs}. 
In the context of scalar quintessence fields it was proposed another 
interesting model with interaction \cite{Huey:2004qv}. It would be worthy 
to mention a model with coupling between dark energy and dark cold matter 
which reproduce power law solutions for energy density \cite{Zimdahl:2002zb}. 
This relation was constrained by the CMB observations \cite{Pavon:2004xk} 
and SNIa observations using statefinder diagnostic parameters (see 
\cite{delCampo:2004wc} and references therein). 

We differ in the presented approach that we do not assume any physical 
mechanism of the energy transfer, which is treated on the phenomenological 
level. We argue that luminosity distance versus redshift relation is the very 
first cosmological test that probes the interactions between dark matter and 
dark energy. This interaction is proposed and is checked whether the ``cosmic 
coincidence conundrum'' is solved. In particular we are interested in the 
direction of energy transfer. For this aim we assume the Friedmann equation 
\begin{equation}
\label{eq:1}
\dot{a}^2 = \frac{\rho}{3} a^2 - k
\end{equation}
where $a$ is the scale factor, $k$ is the curvature index, and a dot means 
the differentiation with respect to the cosmological time $t$. The second 
equation describing the evolution of the model is based on the adiabatic 
condition $T_{\nu;\mu}^{\mu} = 0$ which for the Friedmann-Robertson-Walker 
(FRW) models with some perfect fluid assumes the form 
\begin{equation}
\label{eq:2}
\frac{d}{dt} (\rho a^3) + p \frac{d}{dt} (a^3) = 0.
\end{equation}
Because eq.~(\ref{eq:2}) has the local character the standard interpretation 
is that $\rho$ and $p$ describe the effective energy density and pressure of 
multifluid which do not interact each other. Then eq.~(\ref{eq:2}) describes 
separately the evolution of each component. 

Equation~(\ref{eq:2}) can be rewritten to the form 
\begin{equation}
\label{eq:3}
\dot{\rho} = - 3H(\rho + p)
\end{equation}
where the Hubble function $H=\dot{a}/a$. Let us note that the cosmological 
constant (for which $p=-\rho$) does not contribute to the conservation 
condition~(\ref{eq:3}) as long as it is treated as the noninteracting with 
the rest matter. 

We postulate that apart from dust matter there is dark energy, but both 
fluids interact now and energy can be transported from dark energy sector 
to the nonrelativistic matter sector. Therefore, eq.~(\ref{eq:2}) cannot be 
separable for every component of multifluid. The special case of considered 
class of model are decaying vacuum cosmologies or $\Lambda(t)$ models 
\cite{Bronstein:1933,Lima:2004cq}. The inspiration for constructing the 
noninteracting cosmologies is taken from Wojciulewitsch 
\cite{Wojciulewitsch:1978} (in context of dark energy see 
\cite{Massarotti:1991}).

The main aim of the paper is to show that the observation of distant SNIa 
offer the possibility of testing the energy transport from the vacuum sector 
to the nonrelativistic matter sector which includes dark matter. We show that 
the measurements of third order term in the expansion of the luminosity 
distance relation with respect redshift $z$ (jerk) allows to detect the 
energy transport. Higher order terms in the expansions (snap, crackle, etc.) 
control the velocity, acceleration of energy transport. Note that statefinder 
parameters also control third derivatives but they are inadequate if we want 
to detect the energy transfer directly from observations. 

We assume the different interpretation of eq.~(\ref{eq:2}) rather than its 
modification.

\section{Two-sector models with transfer of energy}

We construct the general class of the decaying dark energy models with the 
interaction starting from the Friedmann first integral which is independent 
of the form of pressure of fluid. We assume for simplicity some two-component 
fluid with effective pressure and energy
\begin{equation}
\label{eq:4}
p_{\text{eff}} = p_{X} + 0, \qquad 
\rho_{\text{eff}} = \rho_{\text{m}} + \rho_{X}
\end{equation}
where $p_{X} = w_{X}\rho_{X}$ ($w_{X} = \text{const}$) describes dark energy 
and $\rho_{\text{m}}$ is the energy of dust matter. If we put $w_{X}=-1$ the 
special case of the cosmological constant is recovered. 

The expression for the conservation condition can be rewritten to the form 
\begin{equation}
\label{eq:5}
\frac{1}{a^3} \frac{d}{dt}(\rho_{\text{m}} a^3) + \frac{1}{a^{3(1+w_{X})}} 
\frac{d}{dt} \left(\rho_{X} a^{3(1+w_{X})}\right) = 0.
\end{equation}
The first term in eq.~\ref{eq:5}) describes the net rate of absorption of 
energy per unit time in unit of comoving volume transfered out of the decaying 
vacuum fluid to the sector of nonrelativistic fluid. If we consider $w=-4/3$ 
the phantom fields are transported. Relation (\ref{eq:5}) is usually 
interpreted without interaction between the sectors. Following Wojciulewitsch 
we postulate that the local energy conservation law (\ref{eq:5}) can be 
written as 
\begin{equation}
\label{eq:6}
\frac{1}{a^3} \frac{d}{dt} \left( \rho_{\text{m}} a^3 \right) = \gamma(t)
\quad \text{and} \quad
\frac{1}{a^{3(1+w_{X})}} \frac{d}{dt} \left(\rho_{X} a^{3(1+w_{X})}\right) = 
- \gamma(t).
\end{equation}
The function $\gamma(t)$ is only a phenomenological description of interaction 
between two sectors. Of course, the exact model of this interaction should 
be taken from the particle physics. If $\gamma(t) > 0$ the energy is transfered 
out of the vacuum, while if $\gamma(t) < 0$ the energy is transfered in the 
opposite direction. 

Integration of eq.~(\ref{eq:6}) gives 
\begin{equation}
\label{eq:7}
\rho_{\text{m}} a^{3} = \rho_{\text{m},0} a_{0}^{3} + \int_{t_{0}}^{t} 
\gamma(t) a^{3} dt 
\quad \text{and} \quad 
\rho_{X} a^{3(1+w_{X})} = \rho_{X,0} a_{0}^{3(1+w_{X})} 
- \int_{t_{0}}^{t} \gamma(t) a^{3(1+w_{X})} dt
\end{equation}
where the index ``0'' means that the quantities are evaluated today. 

It would be useful for our further analysis to represent the Friedmann 
first integral (\ref{eq:1}) in the form for a particle moving in the 
one-dimensional potential
\begin{equation}
\label{eq:8}
\frac{\dot{a}^{2}}{2} + V(a) = \frac{k}{2}, \qquad
V(a) = - \frac{\rho_{\text{eff}} a^{2}}{6} = 
- \frac{\rho_{\text{m}} a^{2}}{6} - \frac{\rho_{X} a^{2}}{6}.
\end{equation}
Because of relation~(\ref{eq:7}) the potential function is explicitly time 
dependent and now takes the following form 
\begin{equation}
\label{eq:9}
V(a) = \frac{1}{2} \left[ \frac{A(t)}{a} + \frac{B(t)}{a^{1+3w_{X}}} \right]
\end{equation}
where
\[
A(t) = \frac{\rho_{\text{m}} a^{3}}{3}, \qquad 
B(t) = \frac{\rho_{X} a^{3(1+w_{X})}}{3}.
\]

In the concordance $\Lambda$CDM models both matter and the cosmological 
constant are treated separately without the interaction, so both functions 
$A(t)$ and $B(t)$ (densities $\rho_{\text{m},0}$ and $\rho_{X,0}$) are 
constant. The presence of the interaction manifests in the model by appearing 
the time dependence of the potential function (\ref{eq:9}). 

Let us note that we postulate the time dependence of $\gamma(t)$ through 
the scale factor, i.e., $\gamma(t)=\gamma(a(t))$, and the potential function 
becomes only a function of $a$ but the exact form of $\gamma(a)$ is required. 
It is convenient to represent the dynamics of the model in terms of the 
Hubble function 
\begin{equation}
\label{eq:10}
H^{2} = \frac{A(t)}{a^{3}} + \frac{B(t)}{a^{3(1+w_{X})}} - \frac{k}{a^{2}} 
= H^{2} (\Omega_{\text{m}} + \Omega_{X} + \Omega_{k}).
\end{equation}
If we postulate that $A(t)=A(a(t))$, $B(t)=B(a(t))$ and put $1+z = a^{-1}$ 
then relation~(\ref{eq:10}) can be used to fit the model parameters 
$\Omega_{i,0}$ to the SNIa data, where $i$ denotes all fluids considered.

Differentiation of both sides of eq.~(\ref{eq:8}) for the potential in the 
form~(\ref{eq:9}) gives the expression for acceleration
\begin{equation}
\label{eq:11}
\ddot{a} = \frac{1}{2} \left[ - \frac{A(a(t))}{a^{2}} 
- \frac{(1+3w_{X})B(a(t))}{a^{2+3w_{X}}} \right] 
\end{equation}
where where we substitute derivatives $\dot{A} = \frac{1}{3} \gamma(t) a^{3}$ 
and $\dot{B} = - \frac{1}{3} a^{3(1+w_{X})} \gamma(t) = - a^{3w_{X}} \dot{A}$ 
from (\ref{eq:7}). Let us rewrite equation (\ref{eq:11}) to the new form
\begin{equation}
\label{eq:12}
q H^{2} = \frac{1}{2} \left[ \frac{A(t)}{a^{3}} + 
\frac{(1+3w_{X})B(a(t))}{a^{3(1+w_{X})}} \right] 
\end{equation}
where $q = - \frac{\ddot{a}}{a H^{2}}$ is the deceleration parameter. 

To control higher derivatives of the scale factor we introduce the 
dimensionless parameter 
\begin{equation}
\label{eq:13}
Q_{n} = (-1)^{n+1} \frac{1}{a H^{n}} \frac{d^{n} a}{dt^{n}}. 
\end{equation}
and then 
\begin{equation}
\label{eq:13a}
Q_{2} = q, \qquad Q_{3} = j, \qquad Q_{4} = s
\end{equation}
are the deceleration, jerk, and snap, respectively \cite{Visser:2003vq}.
In turn, to control the interaction we introduce the dimensionless transfer 
parameter 
\begin{equation}
\label{eq:14}
\nu(t) \equiv \frac{\gamma(t)}{3 H^{3}}
\end{equation}
by analogy to the matter density parameter $\Omega_{\text{m}} = 
\rho_{\text{m}}/3H^{2}$. 

To verify the model we estimate of the parameter $\nu$ at the present 
epoch from the observation of SNIa data. From equations~(\ref{eq:10}) 
and~(\ref{eq:12}) we have 
\begin{equation}
\label{eq:15}
- \Omega_{k} = \frac{k}{a^{2} H^{2}} 
= \frac{3w_{X}}{1+3w_{X}} \Omega_{\text{m}} + \frac{2q}{1+3w_{X}} - 1
\end{equation}
and
\begin{equation}
\label{eq:16}
q = \frac{1}{2} \Omega_{\text{m}} + \frac{1+3w_{X}}{2} \Omega_{X}.
\end{equation}
Of course we have also the constraint condition $\Omega_{k} + \Omega_{\text{m}}
+ \Omega_{X} =1$. 

After the differentiation of both sides of eq.~(\ref{eq:11}) we obtain the 
basic equation relating the jerk to the transfer density parameter 
\begin{equation}
\label{eq:17}
j - \frac{3}{2} w_{X} \nu = \Omega_{\text{m}} + \frac{1}{2} (1+3w_{X}) 
(2+3w_{X}) \Omega_{X}
\end{equation}
and
\begin{equation}
\label{eq:18}
j - \frac{3}{2} w_{X} \nu - 1= \frac{9}{2} w_{X}(w_{X}+1) \Omega_{X} 
- \Omega_{k}.
\end{equation}

Both for strings ($w_{X}=-1/3$) and topological defects ($w_{X}=-2/3$) 
relation~(\ref{eq:17}) does not depend on the density parameter of dark 
energy $\Omega_{X}$. This relation is obvious for all models. 
In the special case of the flat model $\Omega_{X} = 1 - \Omega_{\text{m}}$ 
and then we obtain
\begin{equation}
\label{eq:19}
j - \frac{3}{2} w_{X} \nu = \frac{9}{2} w_{X}(w_{X}+1) \Omega_{X} + 1.
\end{equation}

Summing (\ref{eq:17}) and (\ref{eq:16}) for any $\Omega_{k}$ we obtain 
\begin{equation}
\label{eq:20}
j - \frac{3}{2} w_{X} \nu + q = \frac{3}{2} \Omega_{\text{m}} + 
\frac{3}{2} (1+3w_{X}) (w_{X}+1) \Omega_{X}
\end{equation}
In the special case of the flat model formula~(\ref{eq:20}) reduces to 
\begin{equation}
\label{eq:21}
j - \frac{3}{2} w_{X} \nu + q = - \frac{3}{2} \Omega_{\text{m}} 
(4+3w_{X}) w_{X} + \frac{3}{2} (1+3w_{X})(1+w_{X})
\end{equation}
where $w_{X}$ can be always expressed in terms of $\Omega_{\text{m}}$ and $q$. 
Note that the relation $\gamma(q)$ does not depend on priors on 
$\Omega_{\text{m}}$ for phantoms. 

Finally we obtain that the measurements of the jerk $j_{0}$ at the present 
epoch probes directly the effects of energy transfer as a consequence of 
relation~(\ref{eq:20}). While the cubic term in the relation $d_{L}(z)$ is 
the first term in the Taylor expansion that depends explicitly on the 
$\gamma(t)$, the higher terms in this expansion are related to the derivatives 
of $\gamma(t)$. Let us define the parameter $\nu_{n}$ for the characterization 
of variability of $\gamma(t)$ as
\begin{equation}
\label{eq:22}
\nu_{n} = \frac{1}{3H^{n+3}} \frac{d^{n} \gamma}{dt^{n}}.
\end{equation}
It is obvious that $\nu_{0} = \nu$, $\nu_{1} = \dot{\gamma}/3H^{4}$, \ldots. 

As an illustration that one can control the first derivative of $\gamma(t)$ 
by the measurement of the snap $Q_{4} = s$, we prove the existence of some 
relation obtained by the differentiation of both sides of (\ref{eq:21}). 
To this aim we use the following formulas 
\begin{align}
\frac{dH}{dt} &= - H^{2}(1+q) \\
\frac{d\Omega_{\text{m}}}{dt} &= H[\nu + \Omega_{\text{m}}(2q-1)] \\
\frac{d\Omega_{X}}{dt} &= - H[\nu + \Omega_{X}(1+3w_{X} - 2q)].
\end{align}
Finally we obtain the relation
\begin{multline}
-s + j +3jq + q + 2q^{2} - \frac{3}{2} w_{X} \nu_{1} = 
\frac{3}{2} \nu (3w_{X}^{2} + 4w_{X}) + \frac{3}{2} \Omega_{\text{m}}(2q-1) \\
- \frac{3}{2}(1+3w_{X})(1+w_{X})(1+3w_{X}-2q)\Omega_{X}. 
\label{eq:26}
\end{multline}

Let us briefly comment on the important case of $w_{X}=-1$ corresponding 
the decaying cosmological constant $\Lambda(t)$ cosmologies. Of course, they 
constitute some special case of the considered models
\begin{align}
&j + \frac{3}{2}\nu = 1 - \Omega_{k} \\
&j + \frac{3}{2}\nu + q = \frac{3}{2}\Omega_{\text{m}}
\end{align}

The parameter $\gamma(t) \equiv - d\Lambda/dt$ describes the first derivative 
of $\Lambda$ and therefore the parameter $\nu$ controls its variability. 
In turn the parameter $\nu_{1}$ characterizes the convexity of the function 
$\Lambda(t)$ 
\begin{equation}
\label{eq:28}
-s - j + 3jq + q + 2q^{2} + \frac{3}{2}\nu_{1} + 3\nu + \frac{9}{2}q\nu = 
\frac{3}{2}\Omega_{\text{m}}(2q-1).
\end{equation}

\section{Transfer parameter from distant SNIa}

Let us consider the luminosity distanced versus redshift relation $d_{L}(z)$ 
expanded in the Taylor series with respect to redshift $z$. It can be done 
without knowledge about dynamical equation. For simplicity of presentation 
of the idea of measurement $\nu$, $\nu_1$, \ldots we consider the flat 
universe what is justified by WMAP measurements. Then we obtain 
\cite{Caldwell:2004vi}
\begin{align}
d_{L}(z) &= \frac{z}{H_{0}} \left[ 1 + \frac{1}{2}(1-q_{0})z - 
\frac{1}{6}(1 - q_{0} - 3q_{0}^{2} + j_{0})z^{2} \right] \nonumber \\
\label{eq:29}
&+ \frac{z^{4}}{24H_{0}} \left[ 2 + 2q_{0} - 15q_{0}^{2} - 
15q_{0}^{3} + 5j_{0} + 10q_{0}j_{0} + s_{0} \right] + \cdots 
\end{align}

We propose to detect the time variation of energy transfer using the parameters
$\nu, \nu_{1}, \ldots$. Let us start with estimation of $\nu$ as a first 
approximation. We find the current constraints to the plane $(q_{0},j_{0})$. 
For this aim we mark the shaded region of the $95\%$ confidence level 
constraint from the recent SNIa measurements \cite{Riess:2004nr}. Because 
$\nu = \frac{2}{3}(1-j)$ (or in the general case $\nu = \frac{2}{3} (1-j- 
\Omega_{k,0})$ the detection of the interaction is equivalent to the 
determination whether the jerk is different from $1$ (or $1-\Omega_{k,0}$). 
If $j_{0}<1$ (or $j_{0} < 1-\Omega_{k,0}$) then the energy is transfered from 
the dark energy sector to the nonrelativistic matter sector. If $j_{0}>1$ 
(or $j_{0} > 1-\Omega_{k,0}$) the transport takes place in the opposite 
direction. Note that the negative curvature ($\Omega_{k,0}>0$) makes the 
switch of transfer direction to happen for lower value of $j_{0}$. Therefore, 
to find the direction of transfer we should know not only the value of jerk 
but also the curvature of space. 

In the general case for any $w_{X}$ we have
\begin{equation}
\label{eq:30}
j_{0} = \frac{3}{2} w_{X} \nu + \frac{3}{2} \Omega_{\text{m},0}
\end{equation}
where we use formula (\ref{eq:16}). 

\begin{figure}
\includegraphics[width=0.8\textwidth]{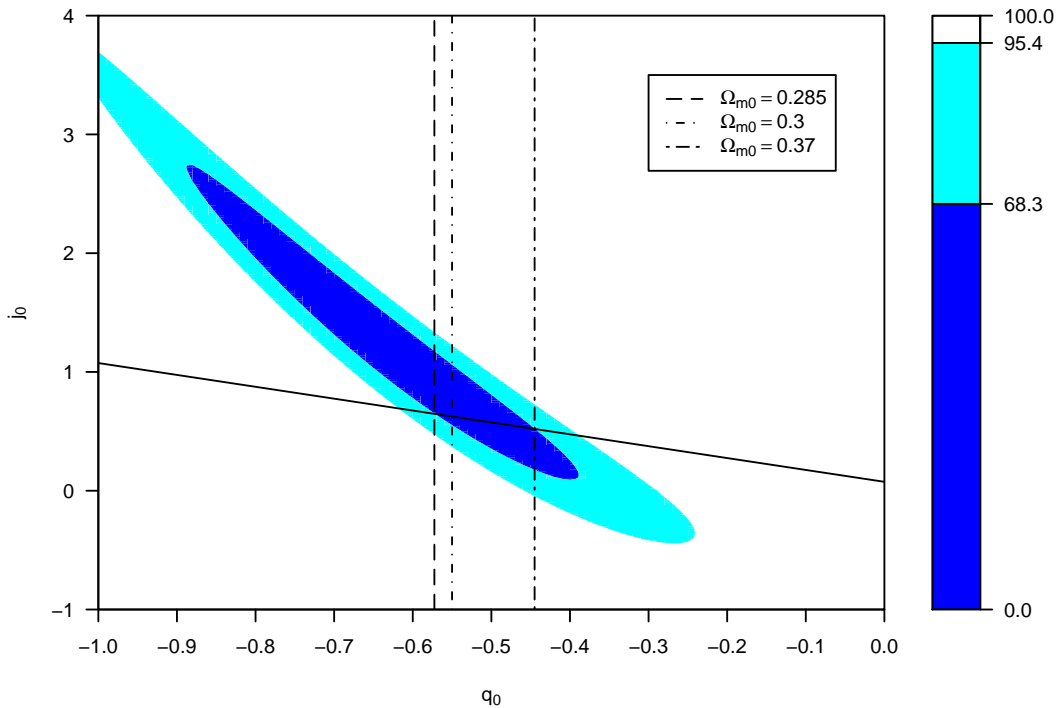}
\caption{The current constraint on the plane 
$(q_{0},j_{0})$. The solid line is the relation $j_{0}(q_{0})$ fulfilled 
in the model. Additionally, for this relation the $1\sigma$ confidence 
level interval for $\Omega_{\text{m},0}$ is drawn.} 
\label{fig:1}
\end{figure}

We consider, for simplicity, the testing of the interaction for the flat 
model and the case $w_{X} =-1$ which corresponds the decaying cosmological 
constant. This allows to substitute $\nu \to w_{X} \nu$. On Fig.~\ref{fig:1}, 
from relation~(\ref{eq:20}) with $w_{X}=-1$, the line 
$j_{0}=q_{0} + \frac{3}{2}(\Omega_{\text{m},0} - \nu)$ is drawn when we 
assume that baryonic matter $\Omega_{\text{m},0} - \nu$ is equal $0.05$. This 
relation allows us to estimate the interval on $\Omega_{\text{m},0}$ and 
$j_{0}$ on the $1\sigma$ confidence level. We mark the line $j_{0}(q_{0})$ 
and the vertical band to denote the interval with the $1\sigma$ confidence 
level for $q_0 \in (-0.5725, -0.445)$ which gives 
$\Omega_{\text{m},0} \in (0.285,0.37)$. In this interval of 
$\Omega_{\text{m},0}$ the jerk $j_0$ is about $0.6$. 

It is very interesting that present SNIa observations allowed us to measure 
the interaction without any special assumptions about physics of the transfer 
process. We thus determined the transfer energy parameter $\nu$ and concluded 
that if we assume that the Universe is flat then the energy transfer takes 
place from the dark energy to dark matter. 

\section{The energy transfer parameter from SNIa data}

\begin{figure}
\includegraphics[width=0.8\textwidth]{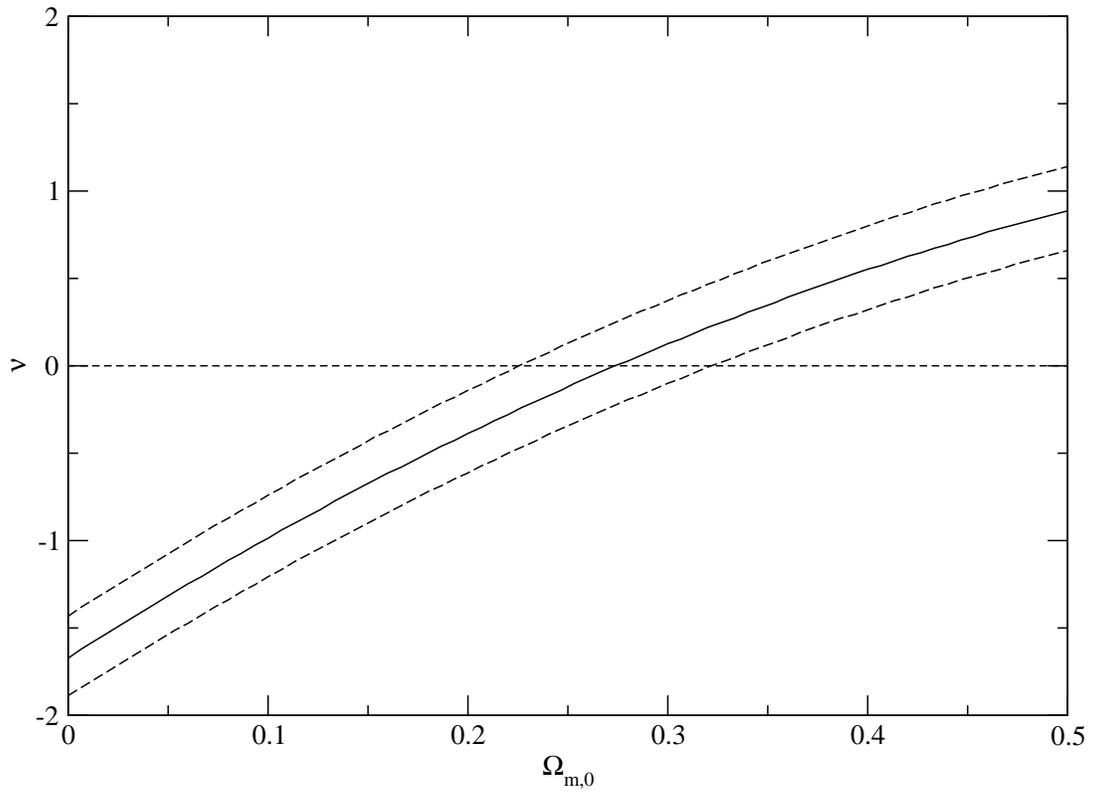}
\caption{The dependency of 
$\Omega_{\text{m},0}$ on $\nu$ for the transfer between decaying vacuum and 
matter sectors.} 
\label{fig:2}
\end{figure}

\begin{figure}
\includegraphics[width=0.8\textwidth]{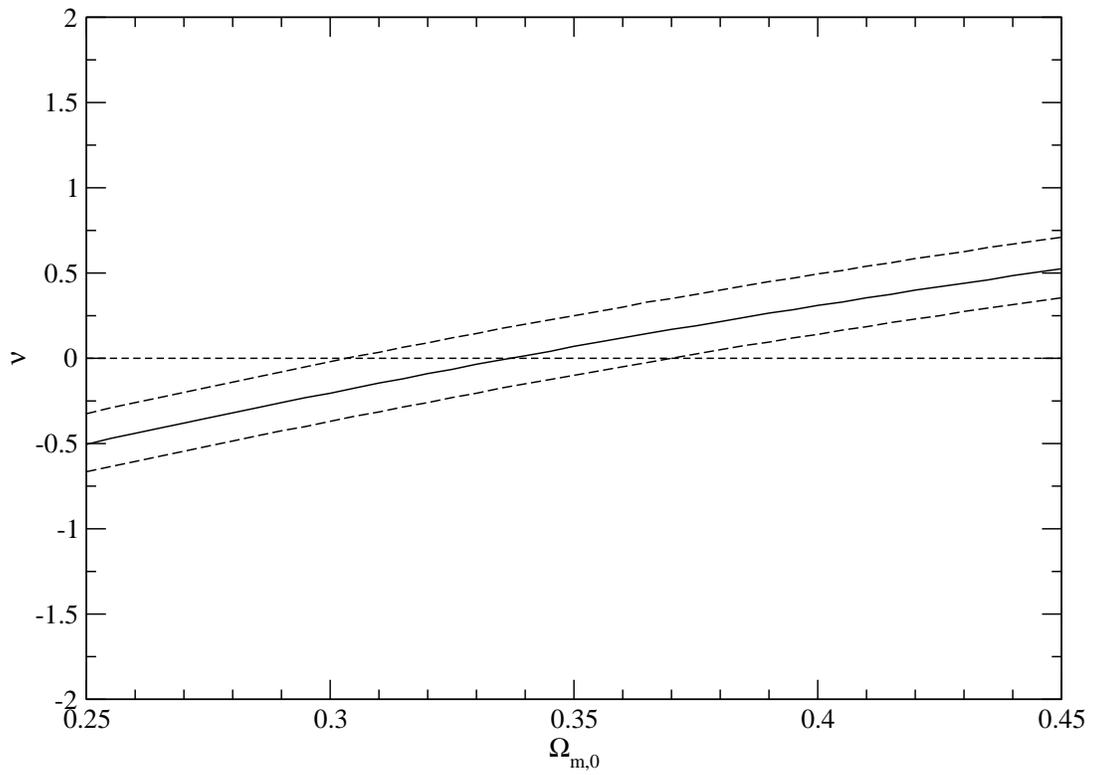}
\caption{The dependency of 
$\Omega_{\text{m},0}$ on $\nu$ for the transfer between phantom and 
matter sectors.} 
\label{fig:3}
\end{figure}

In the previous section it was considered that $\Omega_{\text{b},0} = 
\Omega_{\text{m},0} - \nu$. Now we turn to estimation of the energy 
transfer parameter $\nu$ using the SNIa data. It would be useful to consider 
two situation. First, the energy transfer is between decaying vacuum and 
matter sectors, and second, it is between the phantom ($p_{X}=4/3\rho_{X}$) 
and matter sectors. We would like to answer on two questions. 
\begin{itemize}
\item What is the interval of $\Omega_{\text{m},0}$ which rules out the 
energy energy transfer ($\nu=0$) on the confidence level $95\%$? 
\item Is it possible to tell a scenario with energy transfer and another 
with variable $w_{X}$?
\end{itemize}
To answer to the first question we test the hypothesis that $\nu = 0$.
The transfer from decaying vacuum to matter sectors can be ruled out on 
the confidence level $95\%$ for $\Omega_{\text{m},0} \in (0.23, 0.32)$ 
(Fig.~\ref{fig:2}), while the transfer from phantom to matter sectors is 
ruled out for $\Omega_{\text{m},0} \in (0.30, 0.37)$ on the same confidence 
level (Fig.~\ref{fig:3}). Therefore the transfer between decaying vacuum 
and matter sector seems to be excluded because the extragalactic observations 
and CMB observations favor the values of $\Omega_{\text{m},0}$ in the 
obtained interval. On the other hand these other observations indicate that 
the transfer between phantom and matter sectors is possible. Hence if we 
have other arguments about phantom existence in the universe and if we accept 
that $\Omega_{\text{m},0}\simeq 0.27$ as indicated by WMAP measurements then 
the energy transfer is necessary. 

\begin{figure}
\includegraphics[width=0.8\textwidth]{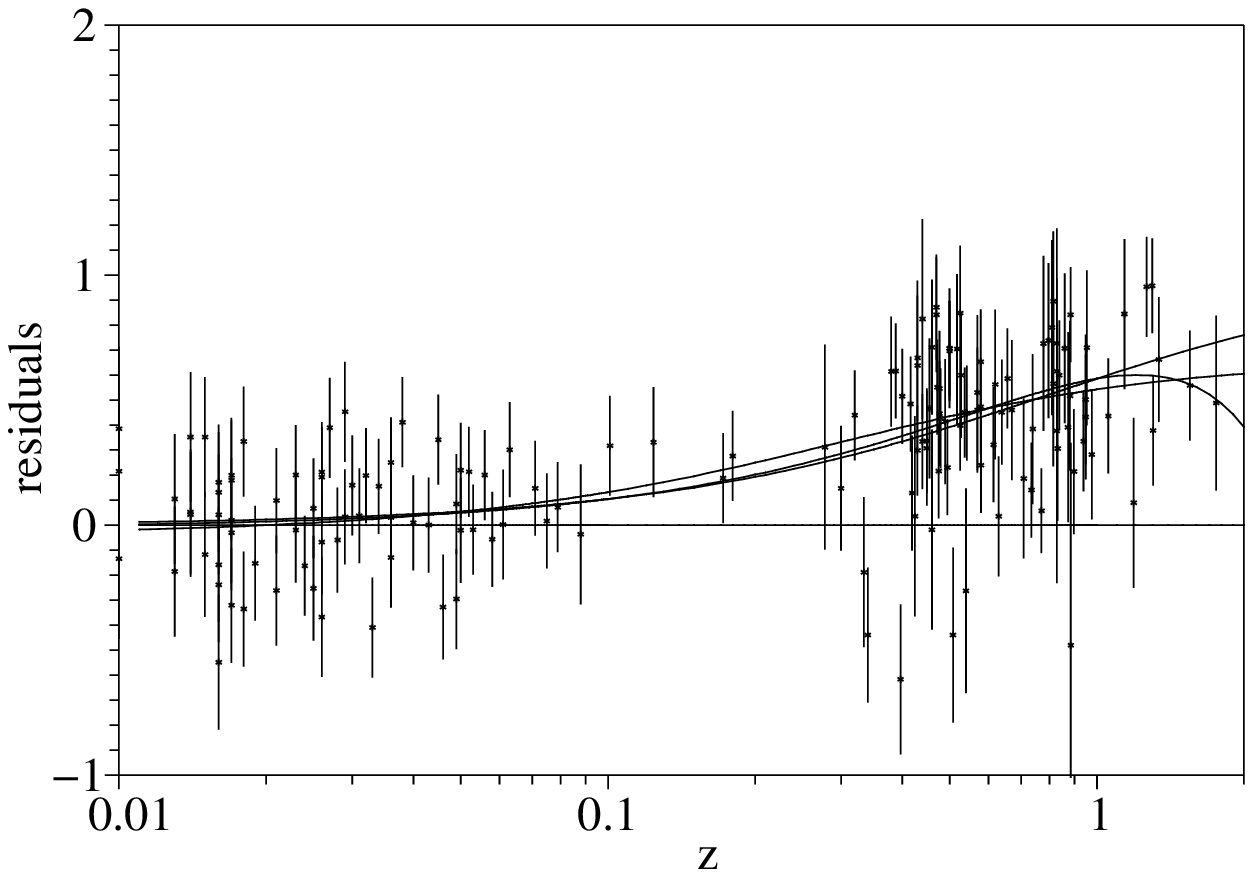}
\caption{The residuals in respect to the Einstein-de Sitter model (the base line) 
for the $\Lambda$CDM model (the upper line), the model with variable $w(z)$ 
(the middle line), and the model with transfer with best fitted $j_0 = 1.26$ 
and $q_0 = -0.64$ (the lower line).} 
\label{fig:4}
\end{figure}

Adopting the same analysis from the previous sections to the case of no energy 
and variable $w(z)=w_{0} + w_{1}z$ we obtain analogous formulas in which 
$\nu w_{0}$ is replaced by $w_1 \Omega_{X,0}$. To answer the second question 
we analyze the Hubble diagram (Fig.~\ref{fig:4}). It is shown that for very 
distant supernovae ($z \simeq 2$) the model with variable $w(z)$ predicts the 
brighter supernovae.

\acknowledgments
The paper was supported by KBN grant no. 1 P03D 003 26. The author is 
very grateful dr A. Krawiec, dr W. God{\l}owski, and T. Stachowiak for 
comments and discussions during the seminar on observational cosmology.


\begin{thebibliography}{22}
\expandafter\ifx\csname natexlab\endcsname\relax\def\natexlab#1{#1}\fi
\expandafter\ifx\csname bibnamefont\endcsname\relax
  \def\bibnamefont#1{#1}\fi
\expandafter\ifx\csname bibfnamefont\endcsname\relax
  \def\bibfnamefont#1{#1}\fi
\expandafter\ifx\csname citenamefont\endcsname\relax
  \def\citenamefont#1{#1}\fi
\expandafter\ifx\csname url\endcsname\relax
  \def\url#1{\texttt{#1}}\fi
\expandafter\ifx\csname urlprefix\endcsname\relax\def\urlprefix{URL }\fi
\providecommand{\bibinfo}[2]{#2}
\providecommand{\eprint}[2][]{\url{#2}}

\bibitem[{\citenamefont{Kuhn}(1962)}]{Kuhn:1962}
\bibinfo{author}{\bibfnamefont{T.~S.} \bibnamefont{Kuhn}},
  \emph{\bibinfo{title}{The Structure of Scientific Revolutions}}
  (\bibinfo{publisher}{University of Chicago Press},
  \bibinfo{address}{Chicago}, \bibinfo{year}{1962}).

\bibitem[{\citenamefont{Perlmutter et~al.}(1999)}]{Perlmutter:1998np}
\bibinfo{author}{\bibfnamefont{S.}~\bibnamefont{Perlmutter}}
  \bibnamefont{et~al.} (\bibinfo{collaboration}{Supernova Cosmology Project}),
  \bibinfo{journal}{Astrophys. J.} \textbf{\bibinfo{volume}{517}},
  \bibinfo{pages}{565} (\bibinfo{year}{1999}), \eprint{astro-ph/9812133}.

\bibitem[{\citenamefont{Riess et~al.}(1998)}]{Riess:1998cb}
\bibinfo{author}{\bibfnamefont{A.~G.} \bibnamefont{Riess}} \bibnamefont{et~al.}
  (\bibinfo{collaboration}{Supernova Search Team}), \bibinfo{journal}{Astron.
  J.} \textbf{\bibinfo{volume}{116}}, \bibinfo{pages}{1009}
  (\bibinfo{year}{1998}), \eprint{astro-ph/9805201}.

\bibitem[{\citenamefont{Freese and Lewis}(2002)}]{Freese:2002sq}
\bibinfo{author}{\bibfnamefont{K.}~\bibnamefont{Freese}} \bibnamefont{and}
  \bibinfo{author}{\bibfnamefont{M.}~\bibnamefont{Lewis}},
  \bibinfo{journal}{Phys. Lett.} \textbf{\bibinfo{volume}{B540}},
  \bibinfo{pages}{1} (\bibinfo{year}{2002}), \eprint{astro-ph/0201229}.

\bibitem[{\citenamefont{Zhu and Fujimoto}(2003)}]{Zhu:2003sq}
\bibinfo{author}{\bibfnamefont{Z.-H.} \bibnamefont{Zhu}} \bibnamefont{and}
  \bibinfo{author}{\bibfnamefont{M.-K.} \bibnamefont{Fujimoto}},
  \bibinfo{journal}{Astrophys. J.} \textbf{\bibinfo{volume}{585}},
  \bibinfo{pages}{52} (\bibinfo{year}{2003}), \eprint{astro-ph/0303021}.

\bibitem[{\citenamefont{Godlowski et~al.}(2004)\citenamefont{Godlowski,
  Szydlowski, and Krawiec}}]{Godlowski:2003pd}
\bibinfo{author}{\bibfnamefont{W.}~\bibnamefont{Godlowski}},
  \bibinfo{author}{\bibfnamefont{M.}~\bibnamefont{Szydlowski}},
  \bibnamefont{and} \bibinfo{author}{\bibfnamefont{A.}~\bibnamefont{Krawiec}},
  \bibinfo{journal}{Astrophys. J.} \textbf{\bibinfo{volume}{605}},
  \bibinfo{pages}{599} (\bibinfo{year}{2004}), \eprint{astro-ph/0309569}.

\bibitem[{\citenamefont{Padmanabhan}(2003)}]{Padmanabhan:2002ji}
\bibinfo{author}{\bibfnamefont{T.}~\bibnamefont{Padmanabhan}},
  \bibinfo{journal}{Phys. Rept.} \textbf{\bibinfo{volume}{380}},
  \bibinfo{pages}{235} (\bibinfo{year}{2003}), \eprint{hep-th/0212290}.

\bibitem[{\citenamefont{Padmanabhan and Choudhury}(2003)}]{Padmanabhan:2002vv}
\bibinfo{author}{\bibfnamefont{T.}~\bibnamefont{Padmanabhan}} \bibnamefont{and}
  \bibinfo{author}{\bibfnamefont{T.~R.} \bibnamefont{Choudhury}},
  \bibinfo{journal}{Mon. Not. Roy. Astron. Soc.}
  \textbf{\bibinfo{volume}{344}}, \bibinfo{pages}{823} (\bibinfo{year}{2003}),
  \eprint{astro-ph/0212573}.

\bibitem[{\citenamefont{Sahni}(2002)}]{Sahni:2002kh}
\bibinfo{author}{\bibfnamefont{V.}~\bibnamefont{Sahni}},
  \bibinfo{journal}{Class. Quantum Grav.} \textbf{\bibinfo{volume}{19}},
  \bibinfo{pages}{3435} (\bibinfo{year}{2002}), \eprint{astro-ph/0202076}.

\bibitem[{\citenamefont{Gorini et~al.}(2004)\citenamefont{Gorini, Kamenshchik,
  Moschella, and Pasquier}}]{Gorini:2004by}
\bibinfo{author}{\bibfnamefont{V.}~\bibnamefont{Gorini}},
  \bibinfo{author}{\bibfnamefont{A.}~\bibnamefont{Kamenshchik}},
  \bibinfo{author}{\bibfnamefont{U.}~\bibnamefont{Moschella}},
  \bibnamefont{and} \bibinfo{author}{\bibfnamefont{V.}~\bibnamefont{Pasquier}}
  (\bibinfo{year}{2004}), \eprint{gr-qc/0403062}.

\bibitem[{\citenamefont{Ziaeepour}(2004)}]{Ziaeepour:2003qs}
\bibinfo{author}{\bibfnamefont{H.}~\bibnamefont{Ziaeepour}},
  \bibinfo{journal}{Phys. Rev.} \textbf{\bibinfo{volume}{D69}},
  \bibinfo{pages}{063512} (\bibinfo{year}{2004}), \eprint{astro-ph/0308515}.

\bibitem[{\citenamefont{Huey and Wandelt}(2004)}]{Huey:2004qv}
\bibinfo{author}{\bibfnamefont{G.}~\bibnamefont{Huey}} \bibnamefont{and}
  \bibinfo{author}{\bibfnamefont{B.~D.} \bibnamefont{Wandelt}}
  (\bibinfo{year}{2004}), \eprint{astro-ph/0407196}.

\bibitem[{\citenamefont{Zimdahl and Pavon}(2003)}]{Zimdahl:2002zb}
\bibinfo{author}{\bibfnamefont{W.}~\bibnamefont{Zimdahl}} \bibnamefont{and}
  \bibinfo{author}{\bibfnamefont{D.}~\bibnamefont{Pavon}},
  \bibinfo{journal}{Gen. Relat. Grav.} \textbf{\bibinfo{volume}{35}},
  \bibinfo{pages}{413} (\bibinfo{year}{2003}), \eprint{astro-ph/0210484}.

\bibitem[{\citenamefont{Pavon et~al.}(2004)\citenamefont{Pavon, Sen, and
  Zimdahl}}]{Pavon:2004xk}
\bibinfo{author}{\bibfnamefont{D.}~\bibnamefont{Pavon}},
  \bibinfo{author}{\bibfnamefont{S.}~\bibnamefont{Sen}}, \bibnamefont{and}
  \bibinfo{author}{\bibfnamefont{W.}~\bibnamefont{Zimdahl}},
  \bibinfo{journal}{JCAP} \textbf{\bibinfo{volume}{0405}}, \bibinfo{pages}{009}
  (\bibinfo{year}{2004}), \eprint{astro-ph/0402067}.

\bibitem[{\citenamefont{del Campo et~al.}(2004)\citenamefont{del Campo,
  Herrera, and Pavon}}]{delCampo:2004wc}
\bibinfo{author}{\bibfnamefont{S.}~\bibnamefont{del Campo}},
  \bibinfo{author}{\bibfnamefont{R.}~\bibnamefont{Herrera}}, \bibnamefont{and}
  \bibinfo{author}{\bibfnamefont{D.}~\bibnamefont{Pavon}},
  \bibinfo{journal}{Phys. Rev.} \textbf{\bibinfo{volume}{D70}},
  \bibinfo{pages}{043540} (\bibinfo{year}{2004}), \eprint{astro-ph/0407047}.

\bibitem[{\citenamefont{Bronstein}(1933)}]{Bronstein:1933}
\bibinfo{author}{\bibfnamefont{M.~V.} \bibnamefont{Bronstein}},
  \bibinfo{journal}{Phys. Z. Sowjetunion} \textbf{\bibinfo{volume}{3}},
  \bibinfo{pages}{73} (\bibinfo{year}{1933}).

\bibitem[{\citenamefont{Lima}(2004)}]{Lima:2004cq}
\bibinfo{author}{\bibfnamefont{J.~A.~S.} \bibnamefont{Lima}},
  \bibinfo{journal}{Braz. J. Phys.} \textbf{\bibinfo{volume}{34}},
  \bibinfo{pages}{194} (\bibinfo{year}{2004}), \eprint{astro-ph/0402109}.

\bibitem[{\citenamefont{Wojciulewitsch}(1978)}]{Wojciulewitsch:1978}
\bibinfo{author}{\bibfnamefont{E.}~\bibnamefont{Wojciulewitsch}},
  \bibinfo{journal}{Acta Cosmologica} \textbf{\bibinfo{volume}{7}},
  \bibinfo{pages}{75} (\bibinfo{year}{1978}).

\bibitem[{\citenamefont{Massarotti}(1991)}]{Massarotti:1991}
\bibinfo{author}{\bibfnamefont{A.}~\bibnamefont{Massarotti}},
  \bibinfo{journal}{Phys. Rev.} \textbf{\bibinfo{volume}{D43}},
  \bibinfo{pages}{346} (\bibinfo{year}{1991}).

\bibitem[{\citenamefont{Visser}(2004)}]{Visser:2003vq}
\bibinfo{author}{\bibfnamefont{M.}~\bibnamefont{Visser}},
  \bibinfo{journal}{Class. Quantum Grav.} \textbf{\bibinfo{volume}{21}},
  \bibinfo{pages}{2603} (\bibinfo{year}{2004}), \eprint{gr-qc/0309109}.

\bibitem[{\citenamefont{Caldwell and Kamionkowski}(2004)}]{Caldwell:2004vi}
\bibinfo{author}{\bibfnamefont{R.~R.} \bibnamefont{Caldwell}} \bibnamefont{and}
  \bibinfo{author}{\bibfnamefont{M.}~\bibnamefont{Kamionkowski}},
  \bibinfo{journal}{JCAP} \textbf{\bibinfo{volume}{0409}}, \bibinfo{pages}{009}
  (\bibinfo{year}{2004}), \eprint{astro-ph/0403003}.

\bibitem[{\citenamefont{Riess et~al.}(2004)}]{Riess:2004nr}
\bibinfo{author}{\bibfnamefont{A.~G.} \bibnamefont{Riess}} \bibnamefont{et~al.}
  (\bibinfo{collaboration}{Supernova Search Team}),
  \bibinfo{journal}{Astrophys. J.} \textbf{\bibinfo{volume}{607}},
  \bibinfo{pages}{665} (\bibinfo{year}{2004}), \eprint{astro-ph/0402512}.

\end{thebibliography}
\end{document}